\documentclass{article}
\usepackage{amsfonts}
\usepackage{amssymb}
\usepackage{amsmath}
\usepackage{amsthm}
\usepackage[normalem]{ulem}
\usepackage[english]{babel}
\usepackage{physics}
\usepackage{cite}
\usepackage[multiple]{footmisc}

\begin{document}

\title{\bf Remarks on Dirac--Bergmann algorithm, Dirac's conjecture and the extended Hamiltonian }

\author{Kirill Russkov${}^{1, 2}$\\
{\small ${}^{1}${\it Department of High Energy and Elementary Particle Physics, Saint Petersburg State University,}}\\ 
{\small {\it Ulianovskaya, 1, Stary Petergof,  Saint Petersburg, Russia}}\\
{\small ${}^{2}${\it Petersburg Nuclear Physics Institute of National Research Centre “Kurchatov Institute”,}}\\
{ \small \it {Gatchina 188300, Russia}} \\
{\small kerill.russ42alex@gmail.com}}
\date{}

\maketitle

\begin{abstract}
The Dirac–-Bergmann algorithm for the Hamiltonian analysis of constrained systems is a nice and powerful tool, widely used for quantization and non-perturbative counting of degrees of freedom. However, certain aspects of its application to systems with first-class constraints are often overlooked in the literature, which is unfortunate, as a naïve treatment leads to incorrect results. In particular, when transitioning from the total to the extended Hamiltonian, the physical information encoded in the constrained modes is lost unless a suitable redefinition of gauge invariant quantities is made. An example of this is electrodynamics, in which the electric field gets an additional contribution to its longitudinal component in the form of the gradient of an arbitrary Lagrange multiplier. Moreover, Dirac’s conjecture, the common claim that all first-class constraints are independent generators of gauge transformations, is somewhat misleading in the standard notion of gauge symmetry used in field theories. At the level of the total Hamiltonian, the true gauge generator is a specific combination of primary and secondary first-class constraints; in general, Dirac’s conjecture holds only in the case of the extended Hamiltonian.

The aim of the paper is primarily pedagogical. We review these issues, providing examples and general arguments. Also, we show that the aforementioned redefinition of gauge invariants within the extended Hamiltonian approach is equivalent to a form of the Stückelberg trick applied to variables that are second-class with respect to the primary constraints.

\end{abstract}

\section{Introduction}

It is well known that the standard Hamiltonian analysis, as taught in most classical mechanics courses worldwide, fails when applied to degenerate Lagrangian systems. This constitutes an issue, since degenerate systems are frequently encountered in physics, most commonly when one is dealing with gauge theories. Indeed, invariance of a theory under a set of local transformations inevitably implies that the Lagrangian is degenerate. This, in turn, leads to restrictions on the initial conditions, referred to as constraints. The constraints complicate the transition from the Lagrangian to the Hamiltonian description. This difficulty was first recognized by Paul Dirac \cite{Dirac50_GenHam, Dirac1951, Dirac58_Royal, Dirac1958b, Dirac1959, DiracLec} in the course of developing a general quantization procedure, and by Peter Bergmann \cite{Bergmann1949, BergmannBrunings1949, BergmannEtAl1950, BergmannGoldberg1955}, whose main motivation was the formulation of a quantum theory of gravity. Independently, they developed an algorithm for dealing with such systems.

Most physicists naturally discuss the Hamiltonian formalism in the context of quantization, and various alternative approaches exist for that purpose \cite{FaddeevJackiw1988, FaddeevSlavnov1980, WeinbergQFT2, Teitelboim}. Our focus here, however, is on Hamiltonian analysis merely as a tool for studying classical theories, for which, in our opinion, the Dirac–Bergmann algorithm provides the most convenient framework. Although it lacks some of the elegance inherent in the Lagrangian formulation, this method is very useful in that it offers a clear iterative procedure: it consists of a sequence of logical steps that reliably leads one to the result\footnote{We recommend this paper\cite{Brown_2022}, which illustrates all the major steps of the algorithm through the example of a single finite-dimensional system.}. The key advantage is that it allows one to rewrite complicated nonlinear systems of second-order equations as first-order equations for canonical pairs.

This work builds on papers \cite{GolovnevHam, GolovnevDegrees, Pons_2005} that discuss subtle and controversial features of the Dirac–Bergmann approach. These features concern the relation between gauge generators and first-class constraints, as well as the extended Hamiltonian, which were introduced in Dirac's \textit{Lectures on Quantum Mechanics}\cite{DiracLec}. We insist that one must exercise care when using this recipe, as these subtleties can actually make the analysis of a system more obscure. The addition of secondary first-class constraints to the Hamiltonian modifies the content of the theory in a peculiar way that keeps the number of dynamical modes intact and makes all first-class constraints independent gauge generators. The latter is not true in the case of the total Hamiltonian, where the correct generators are certain combinations of primary and secondary first-class constraints \cite{Castellani, Pons_2005}. The preservation of the number of dynamical modes is usually mentioned as a justification of Dirac’s extended Hamiltonian. However, constrained modes are not unphysical \cite{GolovnevHam, GolovnevDegrees}. Even though the number of dynamical degrees of freedom remains the same, the system is not equivalent to the one we started with in terms of the original canonical variables. Solutions that describe static configurations become parameterized by arbitrary Lagrange multipliers associated with the secondary first-class constraints in the extended Hamiltonian. Either we can accept this and obtain a strange modification of the original theory, or we can redefine the system by declaring that gauge invariant quantities are those that commute with all first-class constraints \cite{GolovnevDegrees, Pons_2005}. The first option means a loss of static solutions; for example, Coulomb’s law in electrodynamics becomes unphysical. The second is an effective enlargement of the system’s gauge freedom via a reparameterization of variables. It is always attainable, which saves Dirac’s proposals. 

In this paper, we show that the above statements hold for every system with secondary first-class constraints that satisfies reasonable assumptions, including Yang–Mills theory and general relativity. Additionally, it is natural to re-examine some aspects of the Dirac-Bergmann recipe, given the recent realization that it has issues when applied to certain models \cite{GolovnevDegrees, Diracf(Q)_failure}.

\section{A brief reminder of Dirac--Bergmann recipe}

Let us say we have a Lagrangian with $n$ generalized coordinates $\{q_i \}$. In the standard procedure for transitioning to the Hamiltonian formalism, one introduces the generalized momenta $p^i=\frac{\partial L}{\partial \dot{q}_i} (q, \dot{q})$ and then defines the Hamiltonian via the Legendre transform with respect to the velocities, $H=\dot{q}_i p^i -L$\footnote{Repeated indices imply summation.}. Varying the Hamiltonian, together with the requirement of the Lagrange equations, yields Hamilton's canonical equations. This recipe works for the usual models considered in classical mechanics courses because the Hessian of the Lagrangian with respect to velocities, ${\frac{\partial^2 L}{\partial \dot q_i,\partial \dot q_j}}$, is non-degenerate. If degeneracy is present, however, the definitions of the momenta are no longer invertible in terms of velocities, and the above procedure must be modified. The Hessian maps variations of the velocities to variations of the momenta (assuming the coordinates are fixed parameters). If the rank of the Hessian is equal to $n-K$, then only $n-K$ independent variations of velocities are in one-to-one correspondence with $n-K$ independent variations of momenta\footnote{We assume that the mapping $\dot{q} \mapsto p$, given by the definitions of momenta, is regular in some region of $(q,\dot{q})$ values, or in other words, the rank of the Hessian in this region remains constant. This condition ensures that the number of constraints remains constant when transitioning to the canonical description later on. The Dirac-Bergmann approach is based on this crucial assumption.}. Consequently, the same holds for the corresponding velocities and momenta themselves.  We will refer to these as \textit{resolved variables}, and to the remaining ones as \textit{unresolved}. 

The unresolved $K$ velocities are arbitrary free parameters, and their associated momenta are constrained to be functions of other canonical variables. The $K$ constrained momenta impose restrictions on the accessible states of the system. These are called \textit{primary constraints}. Their general form\cite{Castellani, Mukunda1976} can be written as follows
\begin{equation}
\label{eq: constr}
    \phi^a(q,p)=p^a-f^a(q,p^A) =0, \quad a=1, \ldots, K,
\end{equation}
where $\{p^A \}$ denotes the $n-K$ momenta that can be uniquely expressed through the resolved velocities $p^A=f^A(q,\dot q_B,\dot q_a)$. This means our system does not have access to the entire phase space 
$(q,p)$. Rather, it is constrained to a hypersurface defined by (\ref{eq: constr}). 

Performing the Legendre transformation of the Lagrangian $L(q,\dot q_A, \dot q_b)$ with respect to velocities, we obtain
\begin{equation}
 \label{eq: tot Ham vel}
     H(q,p,\dot{q}_b)=f^{-1}_A (q, p^B, \dot q_b) p^A-L(q,f^{-1}_B (q, p^C, \dot q_a), \dot q_b) +\dot q_a  f^a (q,p^B)+\dot{q}_a [ p^a - f^a (q,p^B) ],
\end{equation}
where $f^{-1}_A (q, p^B, \dot q_a)$ is the inverse mapping from $p^A$ to $\dot q_A$, with the variables $(q, \dot q_a)$ treated as fixed parameters. The unresolved velocities remain arbitrary and can therefore be identified with Lagrange multipliers, $\dot{q}_a=\lambda_a$. The resulting Hamiltonian is called \textit{the total Hamiltonian}. It can be rewritten as
\begin{equation}
     \label{eq: tot Ham}
     H_{T}(q,p,\lambda)= H_{can} (q,p^B)+\lambda_a \phi^a (q,p), 
 \end{equation}
 where $H_{can} (q,p^B)=f^{-1}_A (q, p^B, \dot q_b) p^A-L(q,f^{-1}_B (q, p^C, \dot q_a), \dot q_b) +\dot q_a  f^a (q,p^B)$ is \textit{the canonical Hamiltonian} \cite{Teitelboim}. It coincides with the Legendre transform of the Lagrangian with respect to the velocities, with the primary constraints (\ref{eq: constr}) imposed. One can straightforwardly verify that $H_{can}$ does not depend on the unresolved velocities.

 If we consider the variation of both the left and right sides of (\ref{eq: tot Ham}), and equate the coefficients of the identical variations of the arguments, requiring the fulfillment of the Euler-Lagrange equations and the constraints (\ref{eq: constr}), we obtain the equations
 \begin{equation}
    \label{eq: ext ham eq2}
    \dot{q}_a=\{q_a, H_T \}=\lambda_a, \qquad \dot{q}_A=\{q_A, H_{T} \}, \qquad\dot{p}^i=\{p^i, H_{T} \}, \qquad \phi^a (q,p)=0,
\end{equation}
where $\{ \cdot, \cdot \}$ denotes Poisson brackets. The system (\ref{eq: ext ham eq2}) can be viewed as the condition for the stationarity of the \textit{first-order action} $S^{(1)} \equiv \int{dt \ ( \: p^i \: \dot{q}_i -H_{T})}$
which we will use in the future (the action must be varied with respect to both canonical variables and Lagrange multipliers).

\textbf{Classification of constraints.} Constraints in a Hamiltonian system are not limited to just primary ones. To ensure that the primary constraints
\(\phi^a(q(t),p(t))=0\) hold at all times, their time derivatives must vanish:
\(\dot{\phi}^a = \{\phi^a, H_T\} = 0\). These are the \textit{consistency conditions} \cite{Teitelboim}. They may either be satisfied identically, lead to restrictions on the Lagrange multipliers in the form of a system of linear equations, or give rise to new constraints $\chi^u(q,p)=0$\footnote{In his lectures \cite{DiracLec}, Dirac introduces the concepts of weak (\(\approx\)) and strong (\(=\)) equality. A weak equality holds on the hypersurface defined by all constraints, while a strong equality holds on the entire phase space. In this work, we will not introduce a separate notation for equality on the constraint surface.}, which are called \textit{secondary constraints}. The consistency conditions for the secondary constraints may, in turn, give rise to tertiary or higher-order ones. All such subsequent constraints are also referred to as secondary\footnote{Constraints are usually viewed as restrictions on initial data, since they remain satisfied over time if imposed initially. In this work, however, we treat primary constraints as additional equations within the system (\ref{eq: ext ham eq2}). In this approach, secondary constraints follow as consequences of these equations.}. Secondary constraints further reduce the hypersurface on which the system can reside to a hypersurface of smaller dimension.

Another useful classification suggested by Dirac \cite{DiracLec} is the division of constraints into \textit{first-class} and \textit{second-class} constraints.
A function on phase space $f(q,p)$ is called a first-class quantity if it commutes with all constraints under the Poisson brackets on the constraint surface, and second-class if otherwise.

It can be proven that any function that vanishes on the constraint surface can be expressed as a linear combination of all constraints on the entire phase space. Therefore, all first-class constraints form a closed infinite-dimensional algebra under the Poisson brackets: $\{ \Omega^{\rho}(q,p), \Omega^{\sigma} (q,p)\}=C^{\rho  \sigma}{}_{\lambda}(q,p) \: \Omega^{\lambda} (q,p)$,  where $\Omega^{\rho} (q,p)$ denotes both primary and secondary first-class constraints.

\textbf{Gauge symmetry in constrained Hamiltonian systems.} Gauge systems stand out among all Hamiltonian constrained systems since they inevitably possess first-class constraints. The intuition is as follows: requiring a first-class constraint to hold at all times cannot fix any of the Lagrange multipliers in the total Hamiltonian because the Poisson bracket of this constraint with the other constraints in the Hamiltonian vanishes. As a result, some of the Lagrange multipliers remain completely arbitrary, and this arbitrariness in the equations of motion is the gauge freedom: replacing one set of Lagrange multipliers with another in equations (\ref{eq: ext ham eq2}) amounts to a gauge transformation. However, the question of how first-class constraints are related to the generators of such transformations is rather subtle.

The standard answer in the literature is the so-called Dirac's conjecture \cite{DiracLec, Teitelboim}. It states that all first-class constraints, both primary and secondary, are independent generators of gauge transformations. In other words, for any quantity $g(q,p)$, its variation under a gauge transformation has the form
\begin{equation}
    \delta g=\varepsilon_\rho \{g, \Omega^\rho \},
\end{equation}
where $\varepsilon_\rho=\varepsilon_\rho (q,p)$ are independent gauge parameters. 

Following this conjecture, Dirac suggested adding all first-class constraints to the Hamiltonian. The resulting Hamiltonian is called the extended Hamiltonian $H_E = H_{can} + \lambda_{\rho} \Omega^{\rho}$. In Dirac's logic, adding secondary constraints to the Hamiltonian cannot change the physical content since they generate transformations that do not change the physical state. 

\textbf{Remark.} All statements in this section generalize to field theories, with the caveat that spatial coordinates play the role of an index labeling the generalized coordinates. In other words, field theories can be regarded as mechanical systems with an infinite number of degrees of freedom. Consequently, the standard recipe remains valid for well-behaved theories. However, we must note that there are prominent cases when the Dirac-Bergmann algorithm is known to be problematic, such as $f(\mathbb{Q})$ gravity models \cite{Diracf(Q)_failure}, where the consistency conditions yield a system of first-order partial differential equations for the Lagrange multipliers instead of the usual system of linear algebraic equations; the former generally lacks unique solutions. This leads the Dirac-Bergmann algorithm to produce an incorrect number of modes\footnote{An alternative method of mode counting \cite{Heisenberg-f(Q)-Counting} might be of more use in such instances}. Additionally, one can encounter models that possess singular loci in their phase spaces, making the number of dynamical modes ill-defined \cite{GolovnevDegrees}. How to apply the algorithm to such systems is often unclear. We shall put these exotic examples aside in this paper, however, as we are only interested in exploring questions concerning the gauge generators and the extended Hamiltonian in the standard theory of constrained Hamiltonian systems.

\section{A simple toy model}
\textbf{Definitions of modes.} To begin with, let us fix the basic definitions used in this work. We follow the terminology of \cite{GolovnevDegrees}. For the reader’s convenience, we briefly state the relevant definitions below\footnote{The scheme below has certain limitations \cite{GolovnevDegrees}, yet we lack a better one.
}.

\begin{itemize}
    \item We assume that the system's equations of motion are of order not higher than second in time. \textit{\uline{The total number of modes $N_{tot}$}} denotes the total number of real-valued variables in the Lagrangian formulation of the system.

    \item If the time evolution is not uniquely determined by the initial conditions, and there is a freedom in the choice of variables that can be parameterized by arbitrary functions of the spacetime coordinates, we call this freedom \underline{\textit{gauge freedom}}, and the corresponding functions \underline{\textit{gauge parameters}}. We define the number of \underline{\textit{pure gauge modes $N_{gauge}$}} as the number of independent gauge parameters. Among all variables, one can always single out $N_{gauge}$ independent variables whose evolution exhibits this gauge arbitrariness; we will call such variables \underline{\textit{pure gauge}}.

   \item We define the number of \underline{\textit{dynamical modes $N_{dyn}$}} as half of the minimal number of Cauchy data values required to make the evolution of the non-pure gauge variables unique. This definition is motivated by the analogy with classical mechanics: for a particle, one must specify two initial data values: its position and its velocity. The requirement of minimality of $N_{dyn}$ is necessary, since there is often no unambiguous way to decide which variables should be regarded as pure gauge and which should not according to the definition above. We resolve this by identifying as pure gauge those variables that minimize $N_{dyn}$.

\item We define the number of \underline{\textit{constrained modes}} as \underline{\textit{$N_{con}=N_{tot}-N_{dyn}-N_{gauge}$}}. Constrained modes correspond to variables that are uniquely determined by the equations of motion, so no initial conditions needed to be specified for them. We call them physical, since they follow directly from the equations of motion once the pure gauge variables are taken into account.
\end{itemize}

Let us stress one more point. In the common view, constrained modes are often regarded as unphysical. This is usually motivated by the fact that one can always, at least in principle, solve all constraints and express the corresponding variables in terms of the remaining ones. This is true. However, in electrodynamics and gravity, the constrained modes encode static field-configurations -- Coulomb’s and Newton’s laws. Calling them unphysical is therefore a rather odd claim. Nevertheless, regardless of terminology, this physical information is still present in the equations, even after the constraints are solved, and it must be taken into account when analyzing the system.

\textbf{A toy model.} Let us now consider a simple ``toy model''\cite{GolovnevHam, Golovnev-Me} to illustrate our points:
\begin{equation}
    \label{eq: toy model}
    L=\frac{1}{2} (\dot{x}+y)^2.
\end{equation}
Note that the structure of the model is very reminiscent of electrodynamics with coordinate $y$ playing the role of scalar potential -- the Lagrangian does not contain its time derivative.

The Lagrange equations are
\begin{equation}
\label{eq: toy model eqs}
    \ddot{x}+\dot{y}=0, \qquad \dot{x}+y=0.
\end{equation}
As we see, the second equation is a constraint, while the first one directly follows from the second one.

It is evident that the system has a 1-dimensional gauge symmetry under the transformations
\begin{equation}
\label{eq: toy model symmentry}
    x \rightarrow x+\varepsilon(t), \qquad  y \rightarrow y - \dot{\varepsilon} (t) 
\end{equation}
and the quantity $\dot x +y$ is gauge invariant. 

When transitioning to the Hamiltonian formalism, we obtain the momenta $p_x=\dot{x}+y$ and $p_y=0$. Thus, we have a primary constraint $\phi_y \equiv p_y=0$, and the total Hamiltonian of the system is
\begin{equation}
\label{eq: toy model Htot}
    H_{T}=\frac{1}{2} \ p_x^2-y p_x+\lambda_y p_y.
\end{equation}
The equations of motion obtained from this Hamiltonian are
\begin{equation}
\label{eq: toy model eq Htot}
\begin{aligned}
  & \begin{cases}
    \dot{x} = p_x-y \\
    \dot{y} = \lambda_y
  \end{cases}
  \quad \text{and }\quad
  & \begin{cases}
    \dot{p}_x = 0 \\
    \dot{p}_y = p_x
  \end{cases}
\end{aligned}
\quad \text{with the constraint} \quad p_y=0.
\end{equation}
From these equations a secondary constraint $\phi_x\equiv p_x=0$ immediately follows. There are no further secondary constraints, and therefore we have two first-class constraints in total.  Equations (\ref{eq: toy model eq Htot}) reproduce the Lagrange equations (\ref{eq: toy model eqs}), as expected. 

Situations of this type are often referred to as ``gauge hits twice'' or ``gauge strikes twice'' \cite{Teitelboim, GolovnevHam} in the jargon. The peculiarity is that the presence of gauge symmetry leads to two first-class Hamiltonian constraints for every independent symmetry (a primary one and the corresponding secondary one), so that two dynamical modes are eliminated at once: one becomes pure gauge, and the other becomes constrained because of the pair of constraints. In our case, the counting gives $N_{dyn}=0$, $N_{gauge}=1$, $N_{con}=1$.

Note that for this system the ambiguity in specifying Cauchy data, mentioned in the definition above, indeed appears. If we treat \(x\) as a pure gauge variable, we recover the result stated previously. If, instead, we treat \(y\) as a pure gauge variable, we obtain a different counting,
$N_{\text{dyn}}=\tfrac{1}{2}, \  N_{\text{con}}=\tfrac{1}{2}, \ N_{\text{gauge}}=1$.
However, the requirement of minimality of \(N_{\text{dyn}}\) leads to \(N_{\text{dyn}}=0\). This requirement can be justified as follows. If we regard \(y\) as pure gauge, the corresponding gauge-fixing condition is \(y=0\), under which the remaining equation is \(\dot{x}=0\). This appears to give half a dynamical mode. However, a residual gauge freedom with a time-independent parameter (\(\dot{\varepsilon}=0\)) still remains, and it can be used to set this mode to zero. By contrast, the gauge fixing \(x=0\) removes the gauge freedom completely. It is therefore reasonable to attribute this apparent half-mode to the gauge sector.

According to the Dirac conjecture, the generator of gauge transformations has the form $G_D=\varepsilon_y \ p_y+\varepsilon_x \ p_x$, where $\varepsilon_x$, $\varepsilon_y$ are independent gauge parameters. It generates the transformations
\begin{equation}
    \label{eq: toymodel wrong gauge toy model}
    x \rightarrow x+\varepsilon_x, \qquad y \rightarrow y+\varepsilon_y,
\end{equation}
which constitutes a richer gauge symmetry than the transformations (\ref{eq: toy model symmentry}). The two can coincide only under the condition $\varepsilon_y=-\dot{\varepsilon}_x$. This means that first-class constraints do not generate gauge transformations independently of each other. However, one may observe that if gauge transformations are considered only at a fixed instant of time, then the sets of transformations (\ref{eq: toymodel wrong gauge toy model}) and (\ref{eq: toy model symmentry}) coincide. In this limited sense, Dirac's conjecture is correct. Indeed, at a fixed time the gauge parameter and its time derivative can be treated as independent parameters. This ceases to be true if later times are considered. However, this is not the usual notion of gauge symmetry. In field theories, a gauge transformation is a local transformation that does not change the physical state of the system, with parameters depending on all spacetime coordinates.

The reason for this seeming confusion is that Dirac's analysis of gauge generators was incomplete. He considered gauge transformations only in an infinitesimal time neighborhood of the initial conditions. A good review of this can be found in the paper \cite{Pons_2005}. Dirac's logic was as follows \cite{DiracLec}:
Dirac considers an arbitrary canonical variable in a system with first-class constraints and denotes it by $g(t)$. The total Hamiltonian of the system is $H_{T}=\tilde{H}+\lambda_{a} \phi^{a}$, where $\phi^{a}$ are only the primary first-class constraints, and all second-class constraints are included in $\tilde{H}$ together with the canonical Hamiltonian\footnote{Dirac proves that $\tilde{H}$ is a first-class quantity \cite{DiracLec}.}.  
Dirac takes the value of $g$ at the initial instant $t=0$, and then considers the value of $g$ after an infinitesimally small time interval $\delta t$:
\begin{equation}
    \label{eq: Dirac's proof}
    g(\delta t)=g(0)+\delta t( \: \{g, \tilde{H}\}|_{t=0} +\lambda_{a} \{g, \phi^{a}\}|_{t=0}).
\end{equation}
He then proposes to perform an infinitesimal gauge transformation such that the initial value and the transformed value of the variable coincide, $g(0)=g^\prime(0)$, at the initial instant. The value of the variable at time $\delta t$ is then
\begin{equation}
    \label{eq: Dirac proof pt 2}
    g^\prime(\delta t)=g(0)+\delta t( \{g, \tilde{H} \} |_{t=0}+\lambda^\prime_{a} \{g, \phi^{a}\}|_{t=0}).
\end{equation}
Subtracting (\ref{eq: Dirac's proof}) from (\ref{eq: Dirac proof pt 2}), we obtain $\delta g (\delta t)=\varepsilon_{a} \{ g, \phi^{a} \}$, where $\varepsilon_a=\delta t \: \delta \lambda_a$.

Dirac then similarly shows that gauge transformations are generated by quantities of the form $\{\phi^{a}, \tilde{H} \}$ and $\{\phi^{a}, \phi^{b} \}$. He notes that in all examples known to him the secondary first-class constraints have precisely this structure, which prompts him to formulate his famous conjecture. In the end, according to Dirac, transformations that do not change the physical state should be generated by a function
\begin{equation}
\label{eq: Dirac generator}
G_{D}=\varepsilon_{\rho} \: \Omega^{\rho},
\end{equation}
where $\Omega^{\rho}$ are all first-class constraints in the theory and $\varepsilon_{\rho}=\varepsilon_{\rho}(q,p)$ are independent parameters. We will refer to generators (\ref{eq: Dirac generator}) as \textit{\underline{Dirac generators}}.

The issue with this argument is that he performs a first-order perturbative expansion simultaneously in time and in the gauge parameter. As a result, a part of the gauge transformation generated by the secondary constraints is missed. This contribution enters only at higher orders in time. For our toy model, to first order in $\delta t$, Dirac’s reasoning gives the transformations
$x \rightarrow x$ and $y \rightarrow y + \delta t,\delta \lambda$.
However, when one goes to second order, the missing part of the gauge transformation (\ref{eq: toy model symmentry}) is recovered:
$x \rightarrow x - \frac{1}{2}\delta t^2 \delta \lambda$.
Hence, the generators found by Dirac are not valid along the entire system's trajectory in phase space, but rather only for a fixed point on it. \subsection{Castellani Generators}
The problem of finding the correct expression for the generators of gauge transformations in terms of first-class constraints was already posed in the past. A solution was given by Anderson and Bergmann \cite{anderson1951constraints} in 1951 and later refined by Castellani \cite{Castellani} in 1982. The result can be stated as a theorem.

\textbf{Theorem.}  
Suppose there exists an infinitesimal transformation of canonical variables
\[
q \rightarrow q + \delta q, \quad p \rightarrow p + \delta p
\]
such that both the original variables $(q,p)$ and the transformed ones $(q+\delta q, p+\delta p)$ satisfy equations of motion (\ref{eq: ext ham eq2}). Then the generator of this transformation has the form  
\begin{equation}
\label{eq: Castellani generator}
    G = \sum_{k=0}^{m} \varepsilon^{(m-k)}(t) \, G_k,
\end{equation}
where $G_k$ are first-class quantities satisfying the chain of relations  
\begin{equation}
\begin{aligned}
\label{eq: Castellani chain}
    G_0 & = \text{LC of primary constraints}, \\
    G_0 + \{G_1, H_T\} & = \text{LC of primary constraints}, \\
    G_1 + \{G_2, H_T\} & = \text{LC of primary constraints}, \\
    \vdots \\
    G_{m-1} + \{G_m, H_T\} & = \text{LC of primary constraints}, \\
    \{G_m, H_T\} & = \text{LC of primary constraints},
\end{aligned}
\end{equation}
where ``LC'' stands for ``linear combination'', and the notation $\varepsilon^{(m-k)}(t)$ denotes the $(m-k)$-th time derivative of the parameter.

These relations were first obtained by Anderson and Bergmann. Castellani showed that the ``basic generators''\footnote{That is, those which cannot be written as a nontrivial linear combination of other generators.} correspond to chains of minimal length. This minimality is achieved by adding to $G_k$ a linear combination of primary constraints: $G_k \rightarrow G_k + \alpha_a \phi^a.$ We will call the generators obtained in this way \textit{\underline{Castellani generators}} and denote them by $G_C$.

The proof of the theorem is based on applying perturbation theory in the gauge parameters to the equations of motion (\ref{eq: ext ham eq2}) \cite{Castellani, Pons_2005}. Unlike in Dirac's argument, time does not enter the proof as an expansion parameter. To obtain the chains (\ref{eq: Castellani chain}), one considers the variation of the equations (\ref{eq: ext ham eq2}) to first order in the small parameters $\delta q$ and $\delta p$ and requires that all terms distinguishing the transformed equations from the original ones vanish. A thorough analysis of the expressions for local symmetries in terms of constraints can also be found in \cite{Deriglazov:2017:Classical}.

The chain (\ref{eq: Castellani chain}) implies that all $G_k$ are first-class constraints. Moreover, each $G_{k>0}$ includes secondary first-class constraints ($G_k$ is not necessarily a secondary constraint itself, since $G_k$ may also contain a linear combination of primary first-class constraints)\footnote{Nevertheless, in general, the generators do not include all first-class constraints of the theory. An exception arises for constraints of type $\chi^{n}$ and for those that follow from them via the consistency conditions. By constraints of type $\chi^n$, we mean those with the structure $\chi^{n>1}$, where $\chi = 0$ is a regular constraint, i.e., its differential is non-vanishing on the constraint surface.}.

This theorem explains why Dirac generators (\ref{eq: Dirac generator}) produce transformations that do not coincide with gauge transformations. The point is that the generator of a gauge transformation is always a specific combination of primary and corresponding secondary constraints. The coefficients in this combination are not independent, unlike in Dirac's proposal. Consequently, the number of pure gauge modes always coincides with the number of primary first-class constraints, not with the total number of first-class constraints.

If we apply the theorem to our toy model, then the Castellani generator has the form
\begin{equation}
    \label{eq: toy model right generator}
    G_C=-\dot{\varepsilon} p_y+ \varepsilon p_x. 
\end{equation}
It is precisely this generator that produces the correct transformations (\ref{eq: toy model symmentry}).

Let us note once again that the Castellani generator $G_C$ is a gauge symmetry generator in the standard sense --  it generates a local transformation that maps any solution of the equations of motion to another solution describing the same physical state. However, if we restrict ourselves to transformations that preserve the physical state at a fixed instant of time, then Dirac's conjecture does work, since at that instant all time derivatives of the parameter $\varepsilon$ become independent.\footnote{One may notice that Dirac's conjecture is satisfied in the usual notion of gauge symmetry in the trivial case when the system has only primary first-class constraints.}\footnote{An exception, again, is provided by constraints of type $\chi^{n}$. A simple example is the system $L= \frac{1}{2} e^y \dot x^2$ discussed in \cite{Teitelboim}. The system has two first-class constraints, yet only the primary one generates transformations of the spurious variable $y$. At the same time, it is hard to interpret these transformations as a genuine gauge symmetry, since they change the action \cite{GolovnevHam}.} Geometrically, such transformations can be viewed as symmetries of the phase space, foliated into layers -- equivalence classes. Two points belong to the same class if they are related by a transformation generated by the Dirac generator $G_D$. This distinction, which is crucial in our view, is not always made explicit in the literature. As a result, Dirac’s conjecture is often formulated using different underlying notions of gauge symmetry. For example, \cite{conjDirac, Deriglazov:2017:Classical} adopts the standard interpretation, whereas \cite{Teitelboim} considers gauge transformations only at a fixed instant of time. In this paper, we examine the validity of the conjecture in both formulations.

\subsection{On the extended Hamiltonian}

Now let us turn to the extended Hamiltonian formalism.  
For our toy model (\ref{eq: toy model}) the extended Hamiltonian is  
\begin{equation} 
\label{eq: toy model H_E}
H_{E}=\frac{1}{2} \: p^2_x - y \: p_x + \lambda_y \: p_y + \lambda_x \: p_x.  
\end{equation}  
The corresponding equations of motion take the form  
\begin{equation}  
\label{eq: toy model eq Hext}  
\begin{aligned}  
& \begin{cases}  
\dot{x} = p_x - y + \lambda_x \\  
\dot{y} = \lambda_y  
\end{cases}  
\quad \text{and} \quad
& \begin{cases}  
\dot{p}_x = 0 \\  
\dot{p}_y = p_x  
\end{cases} \quad \text{with the constraints} \quad 
& \begin{cases}  
p_x = 0 \\  
p_y = 0. 
\end{cases}  
\end{aligned}  
\end{equation}
The difference from (\ref{eq: toy model eq Htot}) is the appearance of an additional Lagrange multiplier $\lambda_x$ in the first equation in the left column. Because of this term, the system (\ref{eq: toy model eq Hext}) is not equivalent to (\ref{eq: toy model eq Htot}). Indeed, if we eliminate the momenta and write the equations solely in terms of the generalized coordinates, we obtain $\dot{x} + y = \lambda_x$.

At this point there are two possible ways to interpret what happened. A naïve reading is to keep the same interpretation of the variables as before. Then the only physical mode, described by the gauge invariant combination $\dot{x}+y$, seems to lose its physical meaning and become pure gauge, though the dynamical sector has not changed. In that interpretation the mode counting gives $N_{dyn}=0$, $N_{con}=0$, $N_{gauge}=2$. Thus, we have a system with 2 variables and both are unphysical. This makes it look like a system with zero Lagrangian. It is also worth noting that for $p_x$ the standard definition of the momentum, $p_x=\dot x+y$, is no longer satisfied: it is shifted by the new Lagrange multiplier. Imposing this definition forces $\lambda_x$ to vanish, which effectively brings us back to the total Hamiltonian. 

Thus, the extended Hamiltonian system possesses an additional freedom in transforming variables that is absent in the Lagrangian formalism. The gauge transformations take the form  
\begin{equation}  
\label{eq: toy model Hext transformations}  
\delta x = \varepsilon, \quad \text{and} \quad\delta y = -\dot{\varepsilon} + \xi, \quad  \text{where} \quad \xi = \delta \lambda_x, \quad \ddot \varepsilon = -\delta \lambda_y.
\end{equation}  
 One can see that the transformations (\ref{eq: toy model Hext transformations}) coincide with (\ref{eq: toymodel wrong gauge toy model}), since $\xi$ and $\varepsilon$ are independent. Both first-class constraints indeed act as independent gauge generators for the extended Hamiltonian system.

\textbf{Why constrained modes matter. Interaction with constrained modes.}
One might object to our discussion by noting that constrained modes are often treated as redundant, unphysical variables, and it is commonly stated that only dynamical modes carry physical content. If we solve the constraint equation in (\ref{eq: toy model eqs}) for $y$ in terms of $\dot x$ and substitute it back into the Lagrangian, we obtain $L=0$, i.e., an empty system. If both Hamiltonian descriptions have no dynamical modes, then why should the additional gauge freedom of the extended Hamiltonian be a problem? The answer is that, in such situations, the notion that constrained modes are unphysical is not a useful one. Once again, expressing constrained variables through the other ones does not remove the associated physical information. It is simply transferred into the remaining equations of motion and becomes manifest in the presence of couplings. In gauge theories this information encodes the interaction of static configurations of gauge fields with matter fields.

To illustrate this, let us add an interaction to our toy model, à la coupling to fermions in electrodynamics:
\begin{equation}
    \label{eq: toy model interaction}
    L=-\frac{1}{2} \; (\dot x+y)^2 +q^{\ast} (i \: \partial_t - y) q,
\end{equation}
where $q$ is a complex variable. The Lagrangian is invariant under transformations (\ref{eq: toy model symmentry}) together with $q \rightarrow e^{i \varepsilon (t)} q$. The equations of motion are
\begin{equation}
\label{eq: toy model with ferm eqs}
    \ddot{x}+\dot{y}=0, \qquad \dot{x}+y=-q^{\ast} q,  \qquad (i \: \partial_t - y) \: q=0.
\end{equation}
The first two equations imply $\frac{d}{d t} (q^{\ast} q)=0$, i.e., conservation of the charge $q^{\ast} q=C$. If we again solve the second equation for $y$ and substitute it in the Lagrangian, we obtain
$L=q^{\ast} (i \: \partial_t +\dot x) q +\frac{1}{2} (q^{\ast} q)^2$.
Varying the action then yields the same charge conservation and the equation $(i  \: \partial_t +\dot x+q^{\ast} q) q=0$. The residual gauge symmetry allows us to set $x=0$. Using charge conservation, the equation reduces to $(i  \: \partial_t +C)q=0$. Thus the presence of interaction changes the equations for $q$, even though the $(x,y)$-sector still has no dynamical modes. 

When one passes to the extended Hamiltonian, the equation is changed to $(i  \: \partial_t +C+\lambda_x)q=0$. The charge remains conserved, but the arbitrary Lagrange multiplier $\lambda_x$ allows one to absorb its contribution. Now the system can be interpreted as a $L=q^{\ast} (i \: \partial_t) q$, i.e., the $q$-sector without interaction.

\textbf{Correct approach. Redefinition of the Hamiltonian system.} As we have just seen, retaining the original interpretation of the variables leads to an extended Hamiltonian system that is not equivalent to the original one, but instead is a modification in which constrained degrees of freedom are promoted to pure gauge ones.

The structure of the new system, however, suggests a natural way to resolve this mismatch. One must reinterpret the variables. For our toy model, the guiding principle is straightforward: the physically meaningful gauge invariant combination in the Lagrangian picture, $\dot{x}+y=0$, should be identified with the momentum $p_x=0$, as it is gauge invariant under transformations generated by the Dirac generator
$G_D = (-\dot{\varepsilon} + \xi) \: p_y + \varepsilon \: p_y$.
The price is that the set of solutions for the variables $x(t)$ and $y(t)$ in the extended Hamiltonian framework is larger than in the original formulation. This, however, is not problematic, as gauge noninvariant quantities are not observable in any case.

The interpretation of this type was adopted in \cite{conjDirac}, where Dirac's conjecture was claimed to be proven for an arbitrary canonical gauge theory. The key step in their proof was, in fact, to redefine the system so that the conjecture holds: one demands that all first-class constraints generate gauge transformations and, correspondingly, declares as gauge invariant those phase-space functions that commute with all first-class constraints on the constraint surface. Such quantities are usually called canonical gauge invariants\footnote{Canonical gauge invariant quantities are not, in general, the same as gauge invariants in the Lagrangian formalism. For example, the left-hand side of Gau{\ss}'s law in Yang-Mills theory $(D_{i} F^{i0})_a = 0$ is not a scalar with respect to the internal symmetry group, but it is a canonical gauge invariant quantity if expressed through canonical variables (see Section 4).}.

With this definition, the set of canonical gauge invariants is the same in both the total Hamiltonian and the extended Hamiltonian approaches. Indeed, a phase-space function can be invariant under the action of (\ref{eq: Castellani generator}) only if it commutes with all first-class constraints, since $\varepsilon$ is arbitrary. One can further show that the dynamics of these invariants agree with that obtained in the total Hamiltonian approach \cite{conjDirac, Pons_2005}. In electrodynamics, this redefinition means identifying the electric field strength with the spatial momenta and abandoning its standard definition in the longitudinal sector \cite{GolovnevDegrees, Golovnev-Me}.

To see why such a redefinition works, it is useful to view it from a different angle. Let us return to the system with the total Hamiltonian \eqref{eq: toy model Htot}. In this formulation, the gauge symmetry is generated by \eqref{eq: toy model right generator}.
As we have mentioned, the equations \eqref{eq: toy model eq Htot} can be obtained from the first-order Lagrangian
\begin{equation}
\label{eq: L^1_T toy model}
L^{(1)} = p_x \: \dot{x} + p_y \: \dot{y} - \frac{p_x^2}{2} + y \: p_x - \lambda_y \: p_y.
\end{equation}
Suppose now that we want the Dirac conjecture to hold in the strong sense, namely, that the theory should admit an additional independent symmetry generated by the primary constraint, on top of the one already present. In other words, we would like to replace the original generator by $G_D = G_C + \xi \: p_y$, while still preserving equivalence with the initial description. A natural way to achieve this is to make the substitution
$y = y^{\scriptscriptstyle E} - \lambda_x$
directly in the first-order Lagrangian.

This substitution ``splits'' the original variable $y$ into two components and thereby effectively introduces an additional gauge redundancy: the variables $y^{\scriptscriptstyle E}$ and $\lambda_x$ can now be transformed independently. After the substitution, we arrive at
\begin{equation}
\label{eq: L^1_E toy model}
L^{(1)} = p_x \: \dot{x} + p_y \: \dot{y}^{\scriptscriptstyle E} - \frac{p_x^2}{2} + y^{\scriptscriptstyle E} \: p_x - \lambda^{\scriptscriptstyle E}_y \: p_y - \lambda_x \: p_x,
\end{equation}
where $\lambda^{\scriptscriptstyle E}_y = \lambda_y + \dot{\lambda}^{\scriptscriptstyle E}_x$. The first-order Lagrangian \eqref{eq: L^1_E toy model} reproduces precisely the equations of the extended Hamiltonian system \eqref{eq: toy model eq Hext}. Equivalently, one may carry out the substitution directly in the equations \eqref{eq: toy model eq Htot}. The redefined system is equivalent to $L = \frac{1}{2} (\dot{x} + y^{\scriptscriptstyle E} - \lambda_x)^2$. In fact, the procedure we have applied is nothing but the Stückelberg trick. The basic idea is to introduce new variables by reparameterizing the original ones, thereby adding an extra gauge redundancy. A familiar example is an artificial introduction of $U(1)$ gauge invariance for a massive vector field, achieved in essentially the same manner \cite{GolovnevDegrees}. 

The constrained mode is saved, it is now encoded in the momentum $p_x$. This can be formalized as follows. Using our previously introduced prescription for the number of modes, we see that the physical-sector result coincides with that obtained in the total Hamiltonian approach, namely $N_{dyn} = 0$, $N_{con} = 1$, $N_{gauge} = 2$, provided that the new Lagrange multiplier is included in the total number of variables, so that $N_{tot}$ increases to $3$. This is not surprising: we have deliberately enlarged the space of variables by adding an additional pure-gauge degree of freedom $\lambda_x$, which, as shown above, can be introduced already at the Lagrangian level without changing the physical content.

A more general statement can be made: the transition to the extended Hamiltonian framework is always a reparameterization of the first-order Lagrangian. Canonical variables that are second-class with respect to the primary constraint are decomposed into a sum of new canonical variables together with a compensating contribution carried by new Lagrange multipliers, in such a way as to ensure that all first-class constraints act as generators of gauge transformations. This is why one can work with the extended Hamiltonian despite the resulting change in the equations of motion.

\section{Hamiltonian formalism of common gauge field theories}
We now demonstrate that essentially the same line of reasoning applies to two of the most common gauge field theories -- Yang-Mills theory and general relativity. Their structure is, in essence, a more elaborate version of our toy model (\ref{eq: toy model}).

\textbf{A note on mode counting in field theories.} In field theories there is a minor subtlety in defining the number of modes according to the scheme we presented: one must distinguish between local and global degrees of freedom. To illustrate this, let us slightly modify our toy model by introducing one spatial dimension and promoting the variables to fields,
$x=x(\sigma,t)$ and $y=y(\sigma,t)$, while also giving $y$ a spatial derivative. The resulting Lagrangian density is \cite{Golovnev-Me}
\begin{equation}
\label{eq: spatial toy model}
   \mathcal{L} = \frac{1}{2} (\dot x + y^\prime)^2,
\end{equation}
where $y^\prime = \frac{d}{d \sigma} y$. It is noteworthy that this model is equivalent to two-dimensional electrodynamics.
The Euler-Lagrange equations imply that the gauge invariant combination $\dot x + y^\prime$ is constant in both coordinates,
$\dot x + y^\prime = C$. In this case, the dynamical sector is no longer empty: instead of zero dynamical modes, we obtain one half of a dynamical mode due to the integration constant. In total we have: $N_{dyn} = \frac{1}{2}$, $N_{gauge} = 1$, $N_{con} = \frac{1}{2}$. 
However, if we impose vanishing boundary conditions for the gauge invariant quantity $\dot x + y^\prime$, then we get $\dot x + y^\prime = 0$. In this case the familiar result is recovered: $N_{gauge} = 1$, $N_{dyn} = 0$, $N_{con} = 1$. This reflects the fact that the equation $\dot x + y^\prime = C$ can be interpreted either as one half of a global dynamical mode or as zero dynamical modes in a local sense.

From now on, we will always implicitly assume that modes are counted in the local sense, i.e., in the presence of appropriate fall-off boundary conditions for the fields.

\subsection{On canonical formulation of Yang-Mills theory}

Let us begin by considering Yang-Mills theory with an arbitrary $N$-dimensional simple gauge group without sources.
The Lagrangian density of Yang-Mills theory is given by
\begin{equation}
\label{eq: YM L}
    \mathcal{L} = -\frac{1}{4} F^{\mu \nu}_a F^a_{\mu \nu},
\end{equation}
where $F^a_{\mu \nu} = \partial_{\mu} A^a_{\nu} - \partial_{\nu} A^a_{\mu} + g \: C^{a b c} A^b_{\mu} A^c_{\nu}$ is the field strength tensor. $C^{a b c}$ are the structure constants.
The equations of motion of Yang-Mills theory have the form
\begin{equation}
    \label{eq: YM eqs}
    (D_{\mu} F^{\mu \nu})^a = \partial_{\mu} F^{\mu \nu}_a + g \: C^{a b c} A^{b}_{\mu} F^{\mu \nu}_c = 0,
\end{equation}
where $D_{\mu}$ denotes the covariant derivative in the adjoint representation.

The equation for $\nu=0$ is a constraint, since it contains neither time derivatives of $A^a_0$ nor second time derivatives of $A^a_i$. This equation is known as Gau{\ss}'s law. The remaining equations ($\nu=i$) are second-order.

The Yang-Mills fields and the field strength transform under infinitesimal gauge transformations as
\begin{equation}
    \label{eq: YM gauge tr}
    A'^{a}_{\mu} = A^a_{\mu} + g \: C^{a b c} \epsilon^b A^c_{\mu} - \partial_{\mu} \epsilon^a=A^a_{\mu} - (D_{\mu} \epsilon)^a, \qquad F'^{a}_{\mu \nu} = F^{a}_{\mu \nu} + g \: C^{a b c} \epsilon^b F^c_{\mu \nu}.
\end{equation}

From the definition of the generalized momenta $\pi_a^{\mu} \equiv \frac{\delta L}{\delta A^{a}_{\mu}} = F^{\mu 0}_a$, we obtain $N$ primary constraints on the temporal momenta $\pi_a^{0} = 0$ and an explicit expression for the spatial momenta in terms of velocities\footnote{This is not an equality between tensors. It is meant to hold for each component separately.
}
\begin{equation}
\label{eq: YM momenta def}
    \pi^i_a = F_{0 i}^a = \dot A^a_{i} - \partial_{i} A^a_{0} + g \: C^{a b c} A^b_{0} A^c_{i}.
\end{equation}
We see that the variables $A^a_0$ enter the Lagrangian without time derivatives and therefore play the same role as the variable $y$ in the toy model (\ref{eq: toy model}). Consequently, their conjugate momenta vanish.

The total Hamiltonian of the theory is
\begin{equation}
    \label{eq: YM Htot}
    H_T = \int d^3 \vec{x} \biggl( \frac{\pi^i_a{}^2}{2} - (D_i \pi^i)^a \: A^a_0 + \frac{1}{4} F^{a}_{ij}{}^2 + \lambda^a_{0} \: \pi^{0}_a \biggr),
\end{equation}
where integration by parts has been performed and the resulting surface term has been discarded.

The equations of motion take the form
\begin{equation}
    \label{eq: YM ext ham eq}
\begin{aligned}
& \begin{cases}
\dot{A}^a_0 = \lambda^a_{0} \\
F^a_{0i} = \pi^i_a
\end{cases}
\quad \text{and} \quad
\begin{cases}
\dot \pi^0_a = (D_i \pi^i)^a \\
(D_0 \pi^i)^a-(D_k F^{k i})^a = 0
\end{cases}
\quad \text{with the constraints} \quad
\pi^0_a = 0.
\end{aligned}
\end{equation}
From these equations we immediately obtain the secondary constraint $(D_i \pi^i_a) = 0$. All constraints are first-class, as it can be verified by their Poisson bracket algebra,
\begin{equation}
\begin{aligned}
    \label{eq: YM constraint algebra}
    \{\pi^0_a(t,\vec{x}), \pi^0_b(t,\vec{y}) \}=0, \qquad \{\pi^0_a(t,\vec{x}),(D_i \pi^i)^b(t,\vec{y}) \}=0, \qquad  \\
    \{(D_i \pi^i)^a(t,\vec{x}), (D_j \pi^j)^b (t,\vec{y})\}=g \: C^{a b c} (D_k \pi^k)^c(t,\vec{x}) \: \delta^{(3)} (\vec{x}-\vec{y}).
\end{aligned}
\end{equation}
There are no other secondary constraints, since $\{ (D_i \pi^i)^a, H_T\}=-g \: C^{a b c} A^b_0 (D_j \pi^j)^c$. Thus, in total, the system possesses $2N$ constraints. 

As expected, the equation for the time derivative of $A^a_i$ reproduces the definition of the momenta (\ref{eq: YM momenta def}). Substituting this relation into the remaining equations for the momenta yields the Yang-Mills equations of motion (\ref{eq: YM eqs}).

\subsubsection{Generators of gauge transformations}

Again, the Dirac conjecture does not hold for the Hamiltonian formulation of Yang-Mills theory in the standard sense of gauge transformations. To see this, it is sufficient to demonstrate that the Castellani generator produces the gauge group transformations (\ref{eq: YM gauge tr}), whereas the Dirac generator does not. 

The key difference from the toy model's generator (\ref{eq: toy model right generator}) is that the part of the generator corresponding to a zero-order term is not precisely the secondary constraint. It differs from it by a combination of  constraints which one has to add in order to compensate for the contribution $\{ (D_i \pi^i)^a, H_T\}=-g \: C^{a b c} A^b_0 (D_j \pi^j)^c$ in the chain (\ref{eq: Castellani chain}):
\begin{equation*}
    G_0^a(x) = \pi^0_a(x), \qquad G_1^a(x) = (D_i \pi^i)^a(x) + g \: C^{a b c} A^b_0 (D_i \pi^i)^c(x).
\end{equation*}

The Castellani generator can be rewritten in a more compact form
\begin{equation}
    \label{eq: YM Castellani generator}
    G_C = \int d^3 \vec{x} \left( \dot \epsilon_a \: G^a_0 + \epsilon_a \: G^a_1 \right) = \int d^3 \vec{x} \: D_{\mu} \epsilon^a \: \pi^{\mu}_a.
\end{equation}
This generator produces the transformation law (\ref{eq: YM gauge tr}). The fact that this generator yields the correct gauge transformations was shown by Castellani \cite{Castellani}. 

The Dirac generator, in turn, is
\begin{equation}
    \label{YM Dirac genrator}
    G_D = G_C + \int d^3 \vec{x} \: \xi^a \: \pi^{0}_a = \int d^3 \vec{x} \: D_{\mu} \epsilon^a \: \pi^{\mu}_a + \int d^3 \vec{x} \: \xi^a \: \pi^{0}_a,
\end{equation}
where $\epsilon^a$ and $\xi^a$ are independent gauge parameters which precisely means the independence of primary and secondary constraints as generators of symmetries. The generator $G_D$ changes the transformation law form (\ref{eq: YM gauge tr}) to
\begin{equation}
    \label{eq: YM G_D trans}
    \delta A^{a}_{0} = g \: C^{a b c} \epsilon^b A^c_{0} - \dot \epsilon^a + \xi^a, \qquad \delta A^{a}_{i} = g \: C^{a b c} \epsilon^b A^c_{i} - \partial_{i} \epsilon^a, \qquad \delta \pi^{\mu}_a = g \: C^{a b c} \epsilon^b \pi^{\mu}_c.
\end{equation}

The temporal and spatial components of the Yang-Mills field can now be transformed independently of one another, which is a broader symmetry than that of the original system. This, however, is correct in the total Hamiltonian approach only when the physical state is considered at a fixed instant in time.

\subsubsection{The extended Hamiltonians. Constrained and gauge modes}

In total, the theory has $4N$ canonical pairs $(\pi^{\mu}_a, A^b_{\nu})$, together with $N$ primary and $N$ secondary constraints. It also possesses an $N$-dimensional gauge symmetry generated by the Castellani generator. Accordingly, mode counting yields $N_{dyn} = 2N, \ N_{gauge} = N, \ N_{con} = N.$ In other words, ``gauge hits twice'' through the same mechanism as in our toy model. The $N$ constrained modes have a clear physical interpretation: they correspond to Gau{\ss}'s law.
 
Let us now examine how this picture changes upon transition to the extended Hamiltonian formalism,
\begin{equation}
\label{eq: YM Hext}
H_E = \int d^3 \vec{x} \biggl( \frac{\pi^i_a{}^2}{2} - A^a_0 \: (D_i \pi^i)^a + \frac{1}{4} F^{a}_{ij}{}^2 + \lambda^a_0 \: \pi^0_a + \lambda^a_1 \: (D_i \pi^i)^a \biggr).
\end{equation}

From the structure of the extended Hamiltonian, it is evident that, compared to (\ref{eq: YM ext ham eq}), the extended Hamiltonian system's equations of motion are obtained by the replacements
$A^a_0 \to A^a_0 - \lambda^a_{1}$ and $\lambda^a_0 \to \lambda^a_0 - \dot \lambda^a_1$.
They therefore take the form
\begin{equation}
\label{eq: YM H_E ext ham eq}
\begin{aligned}
& \begin{cases}
\dot{A}^a_0 = \lambda^a_{0}, \\
F^a_{0i}+(D_i\lambda_1)^a = \pi^i_a
\end{cases}
\quad \text{and }\quad
\begin{cases}
\dot \pi^0_a = (D_i \pi^i)^a, \\
(D_0 \pi^i)^a-(D_k F^{k i})^a = 0,
\end{cases}
\quad \text{with the constraints} \quad
\pi^0_a = 0.
\end{aligned}
\end{equation}

We encountered exactly the same situation in the model (\ref{eq: toy model H_E}). In particular, we may arrive at the same equations of motion by performing the following reparameterization in the first-order action
$L^{(1)} = \int d^3 \vec{x} \: (\pi^{\mu}_a \dot A_{\mu}^a - H_T)$:
\begin{equation}
\label{eq: YM change of var}
  A^a_0 =  A^{\scriptscriptstyle{E}}{}^a_0-\lambda^a_{1}, \qquad \lambda^a_{0}=\lambda^{\scriptscriptstyle{E}}{}^a_{0}- \dot \lambda^a_{1}.
\end{equation}

The logic is the same as before. If we now include the $N$ new variables $\lambda^a_{1}$ in the total number of modes $N_{tot}$, we obtain
$N_{dyn} = 2N, \ N_{gauge} = 2N, \ N_{con} = N.$
Thus, the number of physical degrees of freedom remains unchanged, while the number of gauge modes increases by a factor of two. In this sense, Dirac's conjecture becomes valid in both interpretations of gauge symmetry transformation. The gauge transformations are now given by (\ref{eq: YM G_D trans}).

At the same time, it is important to emphasize that the variables of the extended Hamiltonian system can no longer be interpreted in the same way as in the original formulation. In particular, we have modified the definition of the chromoelectric fields\footnote{This term is often used in QCD for $F^a_{i 0}$, by analogy with the electric field $F_{i0}$ in electrodynamics.} in terms of the Yang-Mills potentials in the covariantly longitudinal sector\footnote{$F^a_{0i}$ can be decomposed into covariantly longitudinal and covariantly transverse components, by analogy with the decomposition of the electric field in electrodynamics into longitudinal and transverse parts $F^a_{0i}=F^a_{0i}{}^T+F^a_{0i}{}^L$, with $(D_iF_{0i}^T)^a=0$, $(D_iF_{0i}^L)^a\neq0$.}. An analogous situation occurs in electrodynamics \cite{GolovnevHam,Golovnev-Me}. They are now identified not with the components of the field strength tensor $F^a_{0i}$, but rather with the canonical momenta
\begin{equation*}
    \pi^i_a=F^a_{0i}+(D_i \lambda_{1})^a.
\end{equation*}
One may object that the chromoelectric field is not, in any case, an observable since it is not gauge invariant, and therefore, changing its definition should not matter. However, one can construct gauge invariant quantities from it, such as $\text{Tr} \: F^a_{0i}{}^2$, and their relation to the potentials is modified accordingly. More generally, as already noted, what changes is the correspondence between canonical gauge invariant quantities and the variables of the Lagrangian description.

If, nevertheless, one insists that $F^a_{0i}$ is the chromoelectric field, then the physical meaning of Gau{\ss}'s law is lost. Indeed, one finds
\begin{equation}
\label{eq: YM Gauss Law gauge matter}
    (D_i F_{0i})^a=-(D_i D_i \lambda_1)^a, \qquad (D_0 F_{0i})^a - (D_k F_{ki})^a = -D_0 (D_i \lambda_1)^a,
\end{equation}
which implies that a gauge transformation can generate matter fields out of the vacuum. 
Thus, we arrive at a modified Yang-Mills theory rather than an equivalent reformulation. In this modified theory, the additional gauge freedom beyond that of the original system is tied to the freedom to choose an arbitrary configuration of effective charges. The constrained modes become spurious, while the number of dynamical modes remains unchanged -- we lose the covariantly longitudinal component of the chromoelectric field strength.

\subsection{On the ADM formalism of general relativity}

Let us now turn to the canonical formalism of general relativity. A key postulate of GR is the principle of general covariance: all equations of the theory must be written as tensor equations with respect to the group of four-dimensional diffeomorphisms. This group consists of locally invertible smooth transformations of spacetime coordinates,
$x^{\prime \mu}=f^{\mu}(x^{\alpha})$. In particular, the metric transforms as a rank-$2$ tensor under these transformations.

A crucial difference between general relativity and gauge theories such as Yang-Mills theory, is the following: diffeomorphisms, unlike internal symmetry transformations, change not only the functional form of the fields but also the arguments of the fields. Nevertheless, diffeomorphism invariance can be treated formally as an internal symmetry if one considers only the change in the functional dependence of fields. We are therefore interested in the variation of the form $\overline{\delta} g_{\mu \nu} = g^{\prime}_{\mu \nu}(x') - g_{\mu \nu}(x)$, which captures how the metric tensor changes as a function of the local coordinates, rather than how its value changes at a fixed point of the spacetime manifold. This is nothing but the Lie derivative of the tensor field, actually
$\overline{\delta} g_{\mu \nu}=-\mathcal{L}_{\xi} g_{\mu \nu}.$

We will be interested only in infinitesimal coordinate transformations:
\begin{equation}
\label{eq: infinitesimal diffeomorfisms}
    x^{\prime \mu} = x^{\mu} + \xi^{\mu}, \quad |\xi^{\mu}| \sim |\partial_{\alpha} \xi^{\mu}| \ll 1.
\end{equation}
From the standard transformation law of the metric, we obtain the corresponding expression for its variation:
\begin{equation}
    \label{eq: GR metric gauge tr}
    \overline{\delta} g_{\mu \nu} = -(\partial_{\mu} \xi^{\alpha}) g_{\alpha \nu} - (\partial_{\nu} \xi^{\alpha}) g_{\alpha \mu} - \xi^{\alpha} \partial_{\alpha} g_{\mu \nu}, \quad \text{or} \quad
    \overline{\delta} g^{\mu \nu} = (\partial^{\mu} \xi^{\nu}) + (\partial^{\nu} \xi^{\mu}) - \xi^{\alpha} \partial_{\alpha} g^{\mu \nu}.
\end{equation}

Let us consider the gravitational field in the absence of matter. The action of GR is the Einstein--Hilbert action\footnote{We are using the signature $(-,+,+,+)$.}
\begin{equation}
    \label{eq: GR EH Action}
    S_{EH}=\frac{1}{2 \varkappa} \int d^4 x \: \sqrt{-g} \:R.
\end{equation}
Here $\varkappa=8 \pi G$, where $G$ is Newton's gravitational constant. Furthermore, $R=g^{\mu \nu} R_{\mu \nu}$ is the Ricci scalar, $R_{\mu \nu}=R^\alpha{}_{\mu \alpha \nu}$ is the Ricci tensor, and $R^\alpha{}_{\mu \beta \nu}$ is the Riemann--Christoffel curvature tensor.
Varying (\ref{eq: GR EH Action}) yields the famous Einstein field equations,
\begin{equation}
    \label{eq: GR E eqs}
    G^{\mu \nu} \equiv R^{\mu \nu}-\frac{1}{2} \: g^{\mu \nu} R=0.
\end{equation}

Like any gauge theory, GR is a constrained system. It is known that the Einstein tensor can be expressed through the Riemann tensor as
$G^{\mu \nu}=\frac{1}{2} \epsilon^{\mu \rho \sigma \lambda} \epsilon^{\nu \alpha \beta}{}_{\lambda} \:  R_{\rho \sigma \alpha \beta}$,
from which it follows that the components $G^{0 \nu}$ contain only contributions of components $R_{0 i k l}$ and $R_{i k lm}$ and therefore do not involve second time derivatives of the metric. Hence, the equations $G^{0 \nu}=0$ are constraints. The equations $G^{i j}=0$, in turn, are second-order evolution equations.

When passing to the Hamiltonian formalism, one must perform a $(3+1)$ decomposition of the metric, explicitly separating the time coordinate from the spatial ones. Geometrically, this means we foliate spacetime into a one-parameter family of spacelike hypersurfaces $\{t=\text{const}\}$. We implement this decomposition using the ADM variables \cite{ADM, Misner:1973prb}. The line element takes the form
\begin{equation}
    \label{eq: GR ADM-decomp}
    d s^2=g_{\mu \nu} \: dx^{\mu} dx^{\nu}= (-N^2+N_i \: N_j \gamma^{i j}) dt^2+2 N_i dt \: d x^i + \gamma_{i j} dx^i \: dx^j,
\end{equation}
where $\gamma^{i j}$ is the inverse of the spatial metric $\gamma_{i j}$. The functions $N \equiv N_{0}$ and $N_i$ are called the lapse and shift functions, respectively.
Equivalently (\ref{eq: GR ADM-decomp}) can be expressed through the metric components as
$\gamma_{i j}=g_{i j}$, $g_{0 i}=N_i$, and $g_{00}=-N^2+N_i N^i$.
One can then verify that the inverse metric is given by
$g^{00}=-\frac{1}{N^2}$, $g^{0i}=\frac{N^i}{N^2}$, and $g^{ij}=\gamma^{i j}+\frac{g^{0i}g^{0j}}{N^2}$ (we assume $g^{00} \neq 0$).
Finally, from the expansion of the determinant by algebraic complements it follows that
$\sqrt{-g}=N \: \sqrt{\gamma}$.

Let us introduce the unit normal one-form to the hypersurfaces $\{ t=\text{const} \}$ by requiring that $n_{\mu} dx^{\mu}=0$ for any tangent vector $dx^{\mu}$. In terms of the ADM variables, its components are
$n_{\mu}=(-N, \vec{0})$ and $n^{\mu}=(\frac{1}{N},-\frac{N^i}{N})$ \cite{Misner:1973prb, GolovnevADM}.
These objects allow us to define the extrinsic curvature tensor of the hypersurface $t=\text{const}$ as $K_{i j}=-\nabla_i n_j |_{t=\text{const}}=-N \Gamma^0_{i j}$.

From the spatial metric $\gamma_{i j}$, one can construct purely three-dimensional geometric objects. We will denote such three-dimensional quantities by placing a superscript $3$ above the symbol to distinguish them from their four-dimensional counterparts; for instance, $\overset{3}{R}_{i j k l}$, $\overset{3}{R}_{i j }$, and $\overset{3}{R}$.

The following relations hold:
\begin{equation}
\label{eq: ext curv and Gauss eq1}
    K_{i j}=\frac{1}{2 N} \biggl ( \overset{3}{\nabla}_i N_j+\overset{3}{\nabla}_j N_i-\dot \gamma_{i j}  \biggr) \quad \text{and} \quad R_{i k lm}=\overset{3}{R}_{iklm}-K_{im}K_{kl}+K_{il}K_{km}.
\end{equation}

The right equation in (\ref{eq: ext curv and Gauss eq1}) is known as the Gau{\ss} equation in differential geometry. One can show that it leads to the following expression for the Einstein--Hilbert Lagrangian in terms of the ADM variables \cite{GolovnevADM}:
\begin{equation}
    \label{eq: ADM Gauss formula for R}
    \mathcal{L}_{EH}=\sqrt{-g} \: R=\sqrt{\gamma } N \: (\overset{3}{R}+K^{i j} K_{i j}- (K^i_i)^2)-2 \: \partial_0( \sqrt{\gamma} K^i_i)+2 \: \partial_k [\sqrt{\gamma} \: (N^k K^i_i-\gamma^{k l} \partial_l N)].
\end{equation}

We cannot immediately switch to the Hamiltonian formalism in the standard way, starting from the Lagrangian (\ref{eq: ADM Gauss formula for R}), since it contains second time derivatives. For higher-derivative Lagrangians, the transition to the canonical description, in general, requires the Ostrogradsky procedure \cite{Ostrogradsky}. However, one can notice that the second time derivatives appear only in a total derivative term, which can be neglected. Hence, we may start from the Lagrangian
\begin{equation}
    \label{eq: L_ADM}
    L=\frac{1}{2 \varkappa} \int d^3 x \sqrt{\gamma }\: N  (\overset{3}{R}+K^{i j} K_{i j}- (K^i_i)^2).
\end{equation}

\subsubsection{The total Hamiltonian approach}

From the structure of the Lagrangian it is clear that the role of the variable $y$ in the system (\ref{eq: toy model}) is played here by the lapse and shift functions. Indeed, the definition of the momenta yields four primary constraints:
$\pi^0 \equiv \pi_0 \equiv \frac{\delta L}{\delta \dot N} = 0$ and
$\pi^i \equiv \frac{\delta L}{\delta \dot N_i} = 0$.

The remaining momenta, $\pi^{i j} = \frac{\delta L}{\delta \dot{\gamma}_{i j}} = \frac{1}{2 \varkappa} \sqrt{\gamma} (K^l_l \gamma^{i j} - K^{ij})$
can be uniquely expressed in terms of the time derivatives of the spatial metric $\dot \gamma_{i j} = \frac{4 \varkappa}{\sqrt{\gamma}} \: N \biggl( \pi_{i j} - \frac{1}{2} \gamma_{i j} \: \pi^k_k \biggr) + \overset{3}{\nabla}_i N_j + \overset{3}{\nabla}_j N_i$.

After integration by parts and neglecting the surface term, the total Hamiltonian takes the form\footnote{Repeated Greek indices here simply indicate summation and do not imply that the terms transform as tensors. The objects $\mathcal{H}^{\mu}$, $N_{\mu}$, and $\pi^{\mu}$ are not vectors under spacetime diffeomorphisms!}
\begin{equation}
    \label{eq: ADM H_tot}
    H_T = \int d^3 \vec{x} \bigl( N_{\mu} \mathcal{H}^{\mu} + \lambda^{\scriptscriptstyle N}_{\mu} \pi^{\mu} \bigr),
\end{equation}
where\footnote{The momentum $\pi^{k i}$ is a three-dimensional tensor density, so the covariant derivative acts on it not as on an ordinary tensor, but according to the rule $\overset{3}{\nabla}_k {\pi^{k i}} = \sqrt{\gamma} \: \overset{3}{\nabla}_k \bigl(\frac{\pi^{k i}}{\sqrt{\gamma}} \bigr)$.}
\begin{equation}
    \label{eq: ADM H_0}
    \mathcal{H}^0 \equiv \mathcal{H}_0 \equiv -\frac{\sqrt{\gamma}}{2 \varkappa} \: \overset{3}{R} + \frac{1}{2} \: \mathcal{J}_{n m, k l} \: \pi^{n m} \pi^{k l}, \qquad \mathcal{H}^i \equiv -2 \: \overset{3}{\nabla}_k {\pi^{k i}}.
\end{equation}

$\mathcal{J}_{ij,kl}$ is the Wheeler-DeWitt metric. It is given by
$\mathcal{J}_{ij,kl} = \frac{2 \varkappa}{\sqrt{\gamma}} (\gamma_{i k} \gamma_{j l} + \gamma_{i l} \gamma_{j k} - \gamma_{i j} \gamma_{k l})$.
Its inverse has the form
$\mathcal{\hat{J}}^{ij,kl} = \frac{1}{4} \frac{\sqrt{\gamma}}{2 \varkappa} (\gamma^{i k} \gamma^{j l} + \gamma^{i l} \gamma^{j k} - \gamma^{i j} \gamma^{k l})$.

The equations of motion in the total Hamiltonian framework are
\begin{equation}
    \label{eq: ADM N-eqs}
    \dot N_{\mu} = \lambda^{\scriptscriptstyle N}_{\mu}, \qquad \qquad \dot \pi^{\mu} = -\mathcal{H}^{\mu}, \qquad \qquad \pi^{\mu} = 0,
\end{equation}
\begin{equation}
    \label{eq: ADM eqs gamma}
    \dot \gamma_{i j} = \frac{4 \varkappa}{\sqrt{\gamma}} \: N \biggl( \pi_{i j} - \frac{1}{2} \gamma_{i j} \: \pi^k{}_k \biggr) + \overset{3}{\nabla}_i N_j + \overset{3}{\nabla}_j N_i,
\end{equation}
\begin{equation}
    \begin{aligned}
    \label{eq: ADM eqs pi}
    \dot{\pi}^{i j} = -&\frac{N \sqrt{\gamma}}{2 \varkappa} \overset{3}{R}{}^{i j} + \frac{\sqrt{\gamma}}{2 \varkappa} \bigl( \overset{3}{\nabla^i} \overset{3}{\nabla^j} N - \gamma^{i j} \overset{3}{\Delta} N \bigr) +  \frac{4 \varkappa}{\sqrt{\gamma}} \: N \biggl( \frac{1}{2} \pi^{k}{}_k \pi^{i j} - \pi^{i k} \pi_k{}^j \biggr) + \\
    & +N \: \gamma^{i j} \mathcal{H}_0 + \frac{1}{2} N^i \mathcal{H}^j + \frac{1}{2} N^j \mathcal{H}^i - (\overset{3}{\nabla}_k N^i) \pi^{k j} - (\overset{3}{\nabla}_k N^j)\pi^{k i}+ \overset{3}{\nabla}_k (\pi^{i j} N^k).
    \end{aligned}
\end{equation}

The equations (\ref{eq: ADM N-eqs}) show that the system possesses secondary constraints $\mathcal{H}^{\mu} = 0$ -- the gauge, again, ``hits twice''.

One can show that the primary and secondary constraints form a closed algebra:
\begin{equation}
\begin{aligned}
    \label{eq: hypersurf def alg}
    \{\mathcal{H}_i (x), \mathcal{H}_0(x^\prime)\} &= \mathcal{H}_0(x) \: \partial_i \delta^{(3)} (\vec{x} - \vec{x}^\prime), \\
    \{\mathcal{H}_i (x), \mathcal{H}_k(x^\prime)\} &= \mathcal{H}_k(x) \: \partial_i \delta^{(3)} (\vec{x} - \vec{x}^\prime) - \mathcal{H}_i(x^\prime) \: \partial^\prime_i \delta^{(3)} (\vec{x} - \vec{x}^\prime), \\
    \{\mathcal{H}_0 (x), \mathcal{H}_0(x^\prime)\} &= \mathcal{H}^k(x) \: \partial_k \delta^{(3)} (\vec{x} - \vec{x}^\prime) - \mathcal{H}^k(x^\prime) \: \partial^\prime_k \delta^{(3)} (\vec{x} - \vec{x}^\prime).
\end{aligned}
\end{equation}

This proves that all constraints are first-class and that there are no further secondary constraints. Note also that the Hamiltonian vanishes on the constraint surface. This leads to the so-called ``problem of time'' \cite{Problem_of_Time_Anderson}.

The secondary constraints, when expressed through the metric components, are equivalent to the equations $G^{0 \mu} = 0$ since
\begin{equation}
\label{eq: ADM G^0mu(H) eqs}
    G^{00} = -\frac{\varkappa}{2 N^2 \sqrt{\gamma}} \: \mathcal{H}_0, \qquad \qquad
    G^{i 0} = \frac{\varkappa}{2 N \sqrt{\gamma}} \biggl(  \mathcal{H}^i + \frac{N^i}{N} \mathcal{H}^0 \biggr),
\end{equation}
and the equations for the momenta (\ref{eq: ADM eqs pi}) together with (\ref{eq: ADM eqs gamma}) yield the remaining Einstein field equations $G^{i j} = 0$.

\subsubsection{Generators of spacetime diffeomorphisms}

To examine the validity of Dirac's conjecture within the ADM formalism of GR we need to construct the Castellani generator. As in the Yang-Mills case (\ref{eq: YM Castellani generator}), the part of the generator that contains no derivatives of the gauge parameter differs from the secondary constraint by a specific linear combination of constraints with fixed coefficients:
\begin{equation}
    \label{eq: ADM G_C}
    G_{C} = \int d^3 \vec{x} \: \bigl(\epsilon^{\mu} \: G^{(0)}_{\mu} + \dot \epsilon^{\mu} \pi_{\mu}\bigr),
\end{equation}
where
\begin{equation*}
G^{(0)}_0 = \mathcal{H}_0 + \pi_k \partial^k N + \partial_k (N \pi^k) + \partial_k (\pi_0 N), \qquad
G^{(0)}_i = \mathcal{H}_i + (\partial_i N^k) \pi_k + \partial_k (N^k \pi_i) + (\partial_i N) \pi_0.
\end{equation*}

The generator (\ref{eq: ADM G_C}) produces the infinitesimal tensor transformations of the metric induced by infinitesimal spacetime diffeomorphisms. Using the expressions for the metric components in terms of the ADM variables, one can show that
\begin{align}
    \label{eq: ADM gij var}
    \{ g_{i j}, G_C \} & = (\partial_i \xi^{\alpha}) g_{\alpha j} + (\partial_j \xi^{\alpha}) g_{i \alpha} + \xi^{\alpha} \partial_{\alpha} \gamma_{i j} = -\overline{\delta} g_{i j}, \\
    \{ g^{0 \mu}, G_C \} &= -\partial^0 \xi^{\mu} - \partial^{\mu} \xi^0 + \xi^{\alpha} \partial_{\alpha} g^{0 \mu} = -\overline{\delta} g^{0 \mu}, \\
    \text{where} \quad &\xi^{0} = \frac{\epsilon^0}{N}, \qquad \xi^k = \epsilon^k - \epsilon^0 \frac{N^k}{N}.
\end{align}
In the same manner, one can verify that it generates the corresponding transformations for any number of derivatives of the metric, thereby reproducing the full diffeomorphism symmetry of GR.

At the same time, the Dirac generator
\begin{equation}
    \label{eq: ADM G_D}
    G_D = \int d^3 \vec{x} \: \bigl( \epsilon^{\mu} G^{(0)}_{\mu} + (\dot \epsilon^{\mu} + \zeta^{\mu}) \pi_{\mu} \bigr)
\end{equation}
introduces an additional gauge freedom, owing to the extra gauge parameter $\zeta$. It allows one to vary the components $g^{0 \mu}$ independently of the spatial ones:
\begin{equation}
\label{eq: ADM Hext gauge sym}
    \{ g^{00}, G_D \} = \{ g^{00}, G_C \} - 2 \frac{\zeta^0}{N} g^{00}, \qquad \{ g^{0 i}, G_D \} = \{ g^{0 i}, G_C \} - 2 \: \frac{\zeta^0}{N} g^{0 i} +  \zeta^i g^{0 0}.
\end{equation}
Such transformations are not compatible with the general covariance of GR. Thus, the Dirac conjecture again holds only for transformations of the gravitational field's state at a fixed instant. To make it work in a usual sense, we again need to switch to the extended Hamiltonian. 

We have 10 variables in total, 8 constraints, and 4 independent diffeomorphisms \(\xi^{\mu}\). As a result, we find
\(N_{\text{dyn}}=2\), \(N_{\text{gauge}}=4\), and \(N_{\text{con}}=4\). Six of the ten modes are physical, while four correspond to the arbitrariness in the choice of the
coordinate system. The two dynamical modes describe the two polarizations of a gravitational
wave, while the four constrained modes correspond to the four types of metric perturbations around a
background in cosmological perturbation theory: two scalar types and one vector type, the latter
having two independent components.

\subsubsection{The extended Hamiltonian approach}

Upon introducing the extended Hamiltonian, the first-order action and the corresponding equations of motion differ from the total Hamiltonian case by a reparameterization of those variables that are second-class with respect to the primary constraints. In the present case, these variables are the lapse and shift functions:
\begin{equation}
\label{eq: ADM N^E}
    N_{\mu} \equiv N^{\scriptscriptstyle E}_{\mu} + \lambda^{\scriptscriptstyle \mathcal{H}}_{\mu}, \qquad \lambda^{\scriptscriptstyle N}_{\mu} \equiv \lambda^{\scriptscriptstyle E}_{\mu} + \dot \lambda^{\scriptscriptstyle \mathcal{H}}_{\mu}
\end{equation}
The corresponding extended Hamiltonian is
\begin{equation}
\label{eq: ADM H_E}
    H_E = \int d^3 \vec{x} \bigl(  N^{\scriptscriptstyle E}_{\mu} \: \mathcal{H}^{\mu} + \lambda^{\scriptscriptstyle E}_{\mu} \: \pi^{\mu} + \lambda^{\scriptscriptstyle \mathcal{H}}_{\mu} \mathcal{H}^{\mu} \bigr).
\end{equation}
The equations of motion are obtained by substituting (\ref{eq: ADM N^E}) into (\ref{eq: ADM N-eqs}-\ref{eq: ADM eqs pi}).

Let us begin with the correct interpretation of the extended Hamiltonian approach. All objects in the Hamiltonian formulation can be expressed as functionals of the Lagrangian variables -- the spatial metric, lapse, and shifts. The extended Hamiltonian formulation is, once again, equivalent to modifying the functional dependence of the canonical gauge invariants of the theory on the generalized coordinates and velocities. To simplify the analysis, we introduce two types of objects: functionals of the Lagrangian variables arising in the total Hamiltonian formulation and the corresponding functionals evaluated at the extended Hamiltonian's variables $N_{\mu}^{\scriptscriptstyle E}$. We denote objects of the second type with the index $E$, while those of the first type are written without it. For example,
\begin{equation}
\label{eq: ext curv Ext}
    K_{i j}=\frac{1}{2 N} \biggl ( \overset{3}{\nabla}_i N_j+\overset{3}{\nabla}_j N_i-\dot \gamma_{i j}  \biggr) \quad \text{and} \quad K^{\scriptscriptstyle E}_{i j}=\frac{1}{2 N^{\scriptscriptstyle E}} \biggl ( \overset{3}{\nabla}_i N^{\scriptscriptstyle E}_j+\overset{3}{\nabla}_j N^{\scriptscriptstyle E}_i-\dot \gamma_{i j}  \biggr).
\end{equation}
The difference is due to the substitution (\ref{eq: ADM N^E}). Within the correct approach, the standard definitions are no longer satisfied for the spatial momenta. Instead, we have
\begin{equation}
\label{eq: ADM Ext momenta def}
    \pi^{i j} = \Omega \pi_{\scriptscriptstyle E}^{i j} + \Omega L^{i j},
\end{equation}
where
\begin{equation*}
    \Omega = \frac{N^{\scriptscriptstyle E}}{N^{\scriptscriptstyle E} + \lambda^{\scriptscriptstyle \mathcal{H}}_0}, \qquad
    L^{i j} = \frac{\sqrt{\gamma}}{2 \varkappa} \bigl( \Lambda^l_l \: \gamma^{i j} - \Lambda^{i j} \bigr), \qquad
    \Lambda_{i j} = \frac{1}{2 N^{\scriptscriptstyle E}} \bigl( \overset{3}{\nabla}_i \lambda^{\scriptscriptstyle \mathcal{H}}_j + \overset{3}{\nabla}_j \lambda^{\scriptscriptstyle \mathcal{H}}_i \bigr).
\end{equation*}

This immediately leads to modified expressions for the Lagrangian constraints:
\begin{equation}
\label{eq: ADM H_ext neq constr}
    \mathcal{H}_0 = \Omega^2 \: \mathcal{H}^{\scriptscriptstyle E}{}_0 + \Omega^2 \: \tau_0, \qquad \qquad \mathcal{H}^i = \Omega \: \mathcal{H}_{\scriptscriptstyle E}{}^i + \tau^i,
\end{equation}
where
\begin{equation*}
    \tau_{0} = \frac{2 \varkappa}{\sqrt{\gamma}} \bigl( \Lambda^{kl} \Lambda_{k l} - \frac{1}{2} \Lambda^l{}_l{}^2 \bigr) + \frac{2 \varkappa}{\sqrt{\gamma}} \bigl( 2 \pi^{\scriptscriptstyle E}_{k l} \Lambda^{kl} + \pi_{\scriptscriptstyle E}^l{}_l \Lambda^k_k \bigr), \qquad
    \tau^i = - 2 \pi_{\scriptscriptstyle E}^{k i} \partial_k \Omega - 2 \sqrt{\gamma} \overset{3}{\nabla}_k \biggl( \frac{\Omega L^{k i}}{\sqrt{\gamma}} \biggr).
\end{equation*}

The quantities $\mathcal{H}_{\mu}$ serve as the canonical invariants in this formulation and satisfy $\mathcal{H}_{\mu} = 0$, whereas $\mathcal{H}^{\scriptscriptstyle E}{}_{\mu}$ need not vanish on the constraint surface. Accordingly, the correspondence between the Lagrangian and Hamiltonian formalisms is given by (\ref{eq: ADM G^0mu(H) eqs}), rather than by
\begin{equation}
\label{eq: ADM H_E incorrect Einst eqs}
    G^{00} = - \frac{\varkappa}{2 N^{\scriptscriptstyle E}{}^2 \sqrt{\gamma}} \: \mathcal{H}^{\scriptscriptstyle E}{}_0, \qquad \qquad
    G^{i 0} = \frac{\varkappa}{2 N^{\scriptscriptstyle E} \sqrt{\gamma}} \biggl( \mathcal{H}_{\scriptscriptstyle E}^i + \frac{N_{\scriptscriptstyle E}{}^i}{N^{\scriptscriptstyle E}} \mathcal{H}_{\scriptscriptstyle E}^0 \biggr). \qquad
\end{equation}
This redefinition can be viewed as modifying the formulas that relate the metric components to the lapse and shift functions. In particular, we now have
\begin{equation}
\label{eq: ADM H_E metric}
    g^{00} = -\frac{1}{(N^{\scriptscriptstyle E}+\lambda^{\scriptscriptstyle \mathcal{H}}_0){}^2}, \qquad g^{0k} = \frac{N^k+\lambda^{\scriptscriptstyle \mathcal{H}}{}^k}{(N^{\scriptscriptstyle E}+\lambda^{\scriptscriptstyle \mathcal{H}}_0)^2}.
\end{equation}

If we count the new Lagrange multipliers $\lambda^{\scriptscriptstyle \mathcal{H}}_{\mu}$ as part of the total number of modes, then the degree of freedom counting gives $N_{dyn}=2$, $N_{con}=4$, $N_{gauge}=8$. The new Lagrange multipliers $\lambda^{\scriptscriptstyle \mathcal{H}}_{\mu}$ constitute four additional gauge modes that have been introduced into the theory. They provide extra gauge freedom in the form of (\ref{eq: ADM Hext gauge sym}), which is generated by $G_D$.
As in the previous cases, these new gauge modes can also be obtained by making the substitution (\ref{eq: ADM N^E}) directly in the Lagrangian (\ref{eq: L_ADM}).

In the incorrect approach, one assumes that the components of the Einstein tensor are related to the secondary constraints by the formulas (\ref{eq: ADM H_E incorrect Einst eqs}). As in the Yang-Mills theory, this interpretation leads to additional arbitrary matter contributions, in the sense that an effective nonzero energy--momentum tensor appears on the right-hand side of the Einstein equations. From (\ref{eq: ADM H_ext neq constr}) we obtain
\begin{equation}
\label{eq: ADM gauge mattter}
\begin{aligned}
    G^{\mu 0} &= \varkappa \: T^{\mu 0}, \\
    T^{00} = \frac{\tau_0}{2 N^{\scriptscriptstyle E} \sqrt{\gamma}}, \qquad
    &T^{i 0} = -\frac{\tau^i}{2 N^{\scriptscriptstyle E} \sqrt{\gamma} \: \Omega} - \frac{N_{\scriptscriptstyle E}^i}{N^{\scriptscriptstyle E}} \tau_0.
\end{aligned}
\end{equation}
The spatial Einstein equations also receive a nonvanishing contribution as well, but we do not write it out explicitly since the resulting expression is rather cumbersome. The important point is that, in general, all components of this energy-momentum tensor are nonzero. They vanish when $\lambda^{\scriptscriptstyle \mathcal{H}}_{\mu} = 0$, i.e., when the extended Hamiltonian coincides with the total one. The freedom to choose this ``matter'' arbitrarily is precisely what underlies the additional gauge modes. Once again, they make the constrained modes unphysical, and we obtain $N_{dyn}=2$, $N_{con}=0$, $N_{gauge}=8$.

\section{Generalization to all canonical mechanical gauge systems}

We claim that the conclusions drawn from the examples above are universal for all canonical gauge theories in which ``the gauge hits twice''. We will prove this statement for mechanical systems. It turns out that the transition from the total Hamiltonian to the extended Hamiltonian can always be achieved by introducing new spurious variables through reparameterization of the original canonical variables in the first-order action. These new variables subsequently appear as Lagrange multipliers for the secondary constraints in the extended Hamiltonian. This procedure is in complete analogy with what occurs in the toy model, Yang-Mills theory, and the ADM formulation of general relativity. Moreover, it must be applied only to those canonical variables that are second-class with respect to the set of primary constraints.

Suppose we consider a system in which all constraints are first-class. The property that ``the gauge hits twice'' means that to each primary first-class constraint, there corresponds a secondary first-class constraint generated by the requirement of its preservation in time. It is assumed that the system does not have tertiary constraints\footnote{We shall assume that the system has no constraints of the $\chi^n$ type either.}. As mentioned before, the equations of motion for our system follow from the first-order Lagrangian of a constrained system, which has the form
\begin{equation}
    \label{eq: L_T^1 gauge hits twice general form}
    L^{(1)} = \dot q_a p^a + \dot q_A p^A - H_{can}(q, p^B) - \lambda^1_a [p^a - f^a(q, p^B)],
\end{equation}
where we have used the general form of the primary constraints $\phi^a_1 = p^a - f^a(q, p^B), \ a=1 \ldots K$.

One may choose the secondary constraints as
\begin{equation}
    \label{eq: gen arg second constr general form}
    \phi^a_2 (q,p^A) \equiv \{p^a - f^a(q, p^A), H_{can}(q,p^A) \} = -\frac{\partial H_{can}}{\partial q_a} - \frac{\partial f^a}{\partial q_B} \frac{\partial H_{can}}{\partial p^B} + \frac{\partial f^a}{\partial p^i} \frac{\partial H_{can}}{\partial q^i}.
\end{equation}

The form of the reparameterization is motivated by the need for Dirac’s conjecture to hold in the standard interpretation of gauge transformations, as in the examples discussed above. To this end, we must introduce an additional gauge freedom into the dynamics of those canonical variables that are second-class with respect to the primary constraints. This, in turn, ensures an additional independent contribution of primary constraints to the Castellani generator (\ref{eq: Castellani generator}), which gives rise to transformations\footnote{Note that if a variable $z$ commutes with $\phi_1^a$, then the functions $f^a$ do not depend on its conjugate variable, and the corresponding derivative in (\ref{eq: general arg variable trans}) vanishes.}
\begin{equation}
\label{eq: general arg variable trans}
    \delta q_a = \alpha_b \{q_a, \phi_1^b\} = \alpha_a, \quad \delta q_A = \alpha_b \{q_A, \phi_1^b\} = -\alpha_b \: \frac{\partial f^b}{\partial p^A}, \quad \delta p^i = \alpha_b \{p^i, \phi_1^b\} = \alpha_b \: \frac{\partial f^b}{\partial q_i}.
\end{equation}
This extra symmetry can be achieved by the following reparameterization
\begin{equation}
    \label{eq: general arg change of var}
    q_i = q^{\scriptscriptstyle E}_i - \lambda^2_a \{q^{\scriptscriptstyle E}_i, \phi_1^a\}, \qquad p^i = p_{\scriptscriptstyle E}^i - \lambda^2_a \{p_{\scriptscriptstyle E}^i, \phi_1^a\},
\end{equation}
where $\lambda^2_a$ are new Lagrange multipliers. We shall assume that the formulas (\ref{eq: general arg change of var}) are valid to first order in $\lambda^2_a$.  Then the gauge transformation (\ref{eq: general arg variable trans}) corresponds to a shift of the Lagrange multipliers in the extended Hamilton's equations, with $\delta \lambda^2_a = \alpha_a$.

By performing the substitution (\ref{eq: general arg change of var}) into the first-order Lagrangian and keeping terms only up to first order in $\lambda^2_a$, while omitting total derivative terms, we obtain:
\begin{equation}
    \label{eq: general arq L1_E}
    L^{(1)} = p_{\scriptscriptstyle E}^i \dot q^{\scriptscriptstyle E}_i  - H_{can}(q^{\scriptscriptstyle E}, p^A_{\scriptscriptstyle E}) - \lambda_a^{\scriptscriptstyle E}{}^1 \phi^a_1 - \lambda^2_a \phi^a_2,
\end{equation}
which is the first-order Lagrangian of the extended Hamiltonian system. All the contributions that are proportional to the primary constraints are absorbed into the term $\lambda_a^{\scriptscriptstyle E}{}^1 \phi^a_1$. This observation proves that all our arguments apply to any system in which "the gauge hits twice''. Under the correct interpretation of the Extended Hamiltonian the number of physical modes is preserved, but we acquire $K$ new pure gauge modes $\lambda^2_a$.

This statement becomes essentially trivial when the functions $f^a$ depend solely on the coordinates $q_a$. In this case, the functions $f^a$ commute with the canonical Hamiltonian, and the secondary constraints take the form
$\phi^a_2 = -\frac{\partial H_{can}}{\partial q^a} (q, p^A)$. Requiring the constraints $\phi^a_2$ to be first-class implies that they are independent of the unresolved coordinates $q_a$, as they do not depend on the constrained momenta $p_a$ by construction (\ref{eq: gen arg second constr general form}). Consequently, the secondary constraints must enter the canonical Hamiltonian linearly, multiplied by the generalized coordinates that are second-class with respect to them, namely, 
\[
H_{can} = - q_a \, \phi^a_2(q_B, p^A) + \tilde{H}(q_B, p^A).
\]
This structure is precisely what is observed in the toy model (\ref{eq: toy model Htot}), in Yang-Mills theory~(\ref{eq: YM Htot}), and in the ADM formalism~(\ref{eq: ADM H_tot}). Therefore, the substitution (\ref{eq: general arg change of var}) amounts to a shift $- q_a \, \phi^a_2 \rightarrow - (q^{\scriptscriptstyle E}_a+\lambda_a^2) \, \phi^a_2$.

If we express $\phi^a_2$ in terms of the generalized velocities, we obtain Lagrangian constraints, i.e., the Euler--Lagrange equations for the variables $q_a$; since $\dot p^a = -\frac{\partial H_T}{\partial q_a} = \frac{\partial L}{\partial q_a}$ and $\dot f^a = \frac{d}{dt}\frac{\partial L}{\partial \dot q_a}$, the following relation holds
\begin{equation}
    \label{eq: L_constr in terms of H_constr}
    \frac{\delta S}{\delta q_a} = \phi^a_2(q, f^B (q,\dot{q}_A, \dot q_b) ).
\end{equation}
With the incorrect interpretation of the Extended Hamiltonian, however, the newly introduced $K$ pure gauge modes take over the role of the original constrained ones, while the Lagrangian constraints are modified. This occurs because, instead of identifying the Lagrangian constraints with the original functions $\phi^a_2(q, f^B)$, we interpret these constraints as the same functions evaluated at the extended Hamiltonian variables $q^{\scriptscriptstyle E}$ and $\dot q^{\scriptscriptstyle E}$: $\phi^a_2 = \phi^a_2(q^{\scriptscriptstyle E}, f^B (q,\dot{q}^{\scriptscriptstyle E}_A, \dot q^{\scriptscriptstyle E}_b))$. The right-hand sides of the corresponding Euler--Lagrange equations become spurious, as they now depend on the new Lagrange multipliers:
\begin{equation}
    \label{eq: gene arg lag constr}
\frac{\delta S}{\delta q_a} \biggl|_{q=q^{\scriptscriptstyle E}} = \phi^a_2(q^{\scriptscriptstyle E}, f^B (q,\dot{q}^{\scriptscriptstyle E}_A, \dot q^{\scriptscriptstyle E}_b) ) 
=-\frac{\partial \phi^a_2}{\partial q_i}\,
\lambda^2_a \{q^{\scriptscriptstyle E}_i, \phi^a_1\}\Big|_{p^i=\frac{\partial L}{\partial \dot q_i}}
-\frac{\partial \phi^a_2}{\partial \dot q_i}\,
\frac{d}{dt}\!\left(
\lambda^2_a \{q^{\scriptscriptstyle E}_i, \phi^a_1\}\Big|_{p^i=\frac{\partial L}{\partial \dot q_i}}
\right)
\neq 0.
\end{equation}
As we have seen, this can be interpreted as additional matter contributions in Yang-Mills theory (\ref{eq: YM Gauss Law gauge matter}) and in GR (\ref{eq: ADM gauge mattter}). We also find it curious that fixing a gauge in terms of the Lagrange multipliers $\lambda^2_a$ with $\lambda^2_a \neq 0$ leads to a modification of the original system in which the constraints are relaxed \cite{Golovnev-Me}. 

Furthermore, to clarify how the system is modified upon transitioning to the extended Hamiltonian, it is useful to examine the corresponding modification of the Lagrangian. In the case of the total Hamiltonian, the Lagrangian is recovered from the first-order action by substituting the primary constraints and momentum definitions while identifying the unresolved velocities with Lagrange multipliers. A similar procedure can be applied to the extended Hamiltonian by treating the extended Hamiltonian system as equivalent to a total Hamiltonian system in an enlarged phase space $(q^{\scriptscriptstyle E}, p_{\scriptscriptstyle E}, \lambda^2_a, \pi^a)$, described by the action functional $\int dt (p_i \dot{q}^i + \pi^a \dot{\lambda}^{2}_a - H_T-\lambda^{\pi}_a \pi^a)$. Effectively, this approach introduces conjugate momenta $\pi^a$ for the Lagrange multipliers $\lambda^2_a$ that are constrained to vanish on-shell. The total Hamiltonian of this system is given by $H_T+\lambda^{\pi}_a \pi^a$. Yet, the crucial difference here is that the standard velocity-momentum relations are, again, no longer satisfied. Instead, the equations of motion $\dot q^{\scriptscriptstyle E}_A = \{q^{\scriptscriptstyle E}_A, H_E\}$, together with the relation $f_A^{-1}(q^{\scriptscriptstyle E}, p_{\scriptscriptstyle E}^A, \dot q^{\scriptscriptstyle E}_a)=\frac{\partial H_{can}}{\partial p_A}(q^{\scriptscriptstyle E},p^A_{\scriptscriptstyle E})$, lead to $p^A_{\scriptscriptstyle E} = w^A(q^{\scriptscriptstyle E}, \dot q^{\scriptscriptstyle E}, \lambda^2_a)$, where $w^A$ is the solution to the equations
\begin{equation}
\label{eq: ext H new momenta def}
\dot q^{\scriptscriptstyle E}_A - f^{-1}_A (q^{\scriptscriptstyle E}, w^B(q^{\scriptscriptstyle E}, \dot q^{\scriptscriptstyle E}, \lambda^2_a)) + \lambda^2_a \frac{\partial \phi^a_1}{\partial p_{\scriptscriptstyle E}^A} (q^{\scriptscriptstyle E}, w^B(q^{\scriptscriptstyle E}, \dot q^{\scriptscriptstyle E}, \lambda^2_a)) = 0.
\end{equation}
Consequently, the standard Lagrangian is replaced by the following expression:
\begin{equation}
\label{eq: ext lagrangian}
L_{E} = L(q^{\scriptscriptstyle E}, f^{-1}_A(q^{\scriptscriptstyle E}, w^B, \dot q^{\scriptscriptstyle E}_a), \dot q^{\scriptscriptstyle E}_a) + w^A(q^{\scriptscriptstyle E}, \dot q^{\scriptscriptstyle E}, \lambda^2_a) \left(\dot q^{\scriptscriptstyle E}_A - f^{-1}_A (q^{\scriptscriptstyle E}, w^B)\right) - \lambda^2_a \phi^a_1(q^{\scriptscriptstyle E}, w^B).
\end{equation}
The Euler--Lagrange equations derived from $L_E$ are fully equivalent to the equations of the extended Hamiltonian system. This explicitly demonstrates that the dynamics of the original variables are altered -- a conclusion that remains valid for arbitrary $\lambda^2_a$, not only for small perturbations. A detailed treatment of this result is provided in \cite{Deriglazov:2017:Classical}, where $L_E$ is termed the \textit{extended Lagrangian}. In the same work, this Lagrangian is employed as a tool for proving Dirac's conjecture: it is established that all first-class constraints independently generate local symmetry transformations of the extended Lagrangian system.

This discussion generalizes to situations where the gauge ``hits'' arbitrarily many times. Specifically, the tertiary constraints $\phi^a_3$ may arise from the preservation of the secondary constraints $\phi^a_2$, followed by further constraints originating from the consistency conditions of the tertiary ones, and so on, with all the constraints in the resulting chain being first-class\footnote{An example of a system with tertiary constraints $L=\dot x \dot z+yz$ can be found in \cite{Castellani}.}. The reasoning in such cases is analogous: one performs a procedure similar to (\ref{eq: general arg change of var}) that incorporates all first-class constraints in the chain. For example, for a system whose chain ends at the tertiary level, we have:
\begin{equation}
    \label{eq: gen arg chang of var teritary}
    q_i = q^{\scriptscriptstyle E}_i -\lambda^2_a \{q^{\scriptscriptstyle E}_i, \phi_1^a\}- \lambda^3_a \{ q^{\scriptscriptstyle E}_i, \phi^a_2 \}, \qquad p^i = p_{\scriptscriptstyle E}^i - \lambda^2_a \{p_{\scriptscriptstyle E}^i, \phi_1^a\}- \lambda^3_a \{ p_{\scriptscriptstyle E}^i, \phi^a_2 \}.
\end{equation}
The tertiary constraints can be chosen as $\phi^a_3 \equiv \{\phi^a_2, H_{can}\}$. Thus, after performing this substitution and retaining terms only to first order in $\lambda^2_a$ and $\lambda^3_a$, one again arrives at
$L^{(1)} = p_{\scriptscriptstyle E}^i \dot q^{\scriptscriptstyle E}_i - H_E$.

Moreover, the above results extend to systems that possess first-class constraints together with second-class ones. Second--class constraints make it possible to impose restrictions on variables belonging to a single canonically conjugate pair simultaneously and to eliminate these variables from the equations. This elimination can be carried out simply by direct substitution into the total Hamiltonian~\cite{DiracLec}, thereby reducing the system to one with only first-class constraints\footnote{Second--class constraints also correspond to constrained modes. Eliminating the constraints in this manner does not remove the corresponding information they carry from the theory. They must still be accounted for, since they often correspond to observable quantities \cite{GolovnevDegrees}.}. In practice, however, this procedure may turn out to be rather cumbersome.

Thus, we have shown that all our arguments apply to all canonical mechanical gauge theories\footnote{A trivial exception occurs in theories in which the gauge ``hits'' only once. The only possible first-class constraints in such theories are primary ones. In that case the total and extended Hamiltonians coincide, and Dirac's conjecture is automatically satisfied in both versions. An example of this is a relativistic free particle.}. Carrying out an analogous generalization for well-behaved field theories also presents no difficulty.

\section{Conclusions}

In this work, we have explored certain aspects of the Dirac--Bergmann approach to the Hamiltonian formulation of gauge theories that, in our view, do not receive the attention they deserve in the literature. We conclude that the question of gauge symmetry generators for degenerate Hamiltonian systems, and the treatment of such systems within the extended Hamiltonian framework, are more subtle than they may appear. Ignoring these nuances leads to a naïve interpretation in which the canonical variables are assumed to play the same role as in the total Hamiltonian approach. In this case, the system appears to possess twice the actual gauge symmetry, since Dirac's conjecture is fulfilled, with the constrained modes becoming spurious.

Although there is nothing inherently wrong with introducing additional gauge freedom into a theory, and even though such a naïve interpretation leaves the dynamical modes intact, it is essential to account for the constrained part of the physical sector, as it corresponds to non-dynamical observables. Constrained modes must therefore be expressed in terms of combinations that commute with the first
-class constraints in the extended Hamiltonian approach. This actually implies that we must redefine the system's observables. Interestingly, this redefinition can be interpreted as a version of the Stückelberg trick applied to variables that are second-class with respect to the set of all primary first-class constraints.

\section*{Conflicts of Interest} The author declares no conflicts of interest.

\section*{Acknowledgements}
The author thanks the organizers of the Eighth International Conference “Models in Quantum Field Theory”, dedicated to Alexander Nikolaevich Vasiliev, and is also grateful to Alexey Golovnev and Sergey Paston for useful discussions.

\bibliographystyle{unsrt}
\bibliography{sample}

\end{document}